\documentclass[prd,showpacs,twocolumn,superscriptaddress,floatfix,nofootinbib,10pt]{revtex4-2}
\usepackage{setspace}
\usepackage[utf8]{inputenc}
\usepackage{float}
\usepackage{amsmath,amssymb,amsfonts,bm}
\usepackage{graphicx}
\usepackage{multirow}
\usepackage[usenames,dvipsnames]{color}
\allowdisplaybreaks
\usepackage[
colorlinks=true,
linkcolor=blue,
breaklinks=true,
urlcolor=blue,
citecolor=blue]{hyperref}

\usepackage{orcidlink}
\usepackage{subfigure}
\usepackage{soul}
\usepackage{xcolor}
\usepackage{physics}
\usepackage{booktabs}
\usepackage{ulem}

\definecolor{green}{rgb}{0,0.6,0}
\newcommand{\mev}{\textrm{ MeV}}

\newcommand{\GXNU}{\affiliation{Department of Physics, Guangxi Normal University, Guilin 541004, China}}

\newcommand{\GXZD}{\affiliation{Guangxi Key Laboratory of Nuclear Physics and Technology, Guangxi Normal University, Guilin 541004, China}}

\newcommand{\IFIC}{\affiliation{Departamento de F\'{\i}sica Te\'orica and IFIC, Centro Mixto Universidad de
Valencia-CSIC Institutos de Investigaci\'on de Paterna, Apartado 22085,
46071 Valencia, Spain}}

\begin{document}
\title{The $B^{+(0)} \to \bar D^{0(-)} D^{*}_{s0}(2317)^+$ decays and the molecular structure of $D^*_{s0}(2317)$}

\begin{abstract}
We have conducted a study of the $B^{+(0)} \to \bar D^{0(-)} D^{*}_{s0}(2317)^+$ reactions from the perspective that the $D^*_{s0}(2317)$ resonance is a molecular state of the $DK$ and $D_s \eta$ components. We have followed a method to evaluate the branching fractions obtaining information from the experimental data on the $B^+\to \bar D^0 K^+ D^0$, $B^+\to \bar D^0 K^0 D^+$, $B^0 \to D^- K^+ D^0$, $B^0 \to D^- K^0 D^+$ reactions, which have the $D^0 K^+$ and $D^+ K^0$ pairs in the final state. The approach concentrates the dynamics of the weak process in the branching ratios of these reactions and pays attention to the propagation of the $DK$ components and their strong interaction to form the $D^*_{s0}(2317)$ resonance. By means of two free parameters, we are able to describe these six rates, showing consistency with the molecular picture of the $D^*_{s0}(2317)$ state.
\end{abstract} 

\author{Wei-Hong Liang\orcidlink{0000-0001-5847-2498}}%
\email{liangwh@gxnu.edu.cn}
\GXNU%
\GXZD%

\author{Zhuo-Ran Hu}
\GXNU%

\author{Eulogio Oset\orcidlink{0000-0002-4462-7919}}%
\email{Oset@ific.uv.es}
\GXNU%
\IFIC%

\maketitle

\section{Introduction}\label{sec:Intr}
The structure of the $D^*_{s0}(2317)$ state has been an object of debate for some time.
There are many works assuming a standard $s\bar q$ nature \cite{Godfrey:1985xj,Ebert:1997nk,Godfrey:2003kg,Ishida:2003gu,Close:2005se,Wang:2006mf,Wei:2005ag,Colangelo:2003vg,Goity:2000dk,Liu:2006jx,Wang:2006fg}.
The chiral doublet model has also been advocated in Ref.~\cite{Bardeen:2003kt}.
On the other hand, the existence of large molecular components of $DK, D_s\eta$ has also been advocated in other works \cite{vanBeveren:2003kd,Barnes:2003dj,Chen:2004dy,Kolomeitsev:2003ac,Gamermann:2006nm,Guo:2006rp,Yang:2021tvc,Liu:2022dmm,Faessler:2007gv,Su:2025aiz}.
This latter nature of the $D^*_{s0}(2317)$ has also received support from Lattice QCD calculations \cite{Mohler:2013rwa,Lang:2014yfa,Bali:2017pdv,Cheung:2020mql,MartinezTorres:2014kpc},
and in Ref.~\cite{MartinezTorres:2014kpc} the $DK$ component probability was determined at the level of $72\%$.
Further support for this molecular structure is obtained from the analysis of $B_s \to \pi^+ \bar D^0 K^-$, $B^0 \to D^- D^0 K^+$ decays \cite{BaBar:2014jjr,LHCb:2014ioa},
where the $\bar D \bar K, DK$ mass distributions at threshold are very much enhanced by the presence of the $D^*_{s0}(2317)$ state below threshold \cite{Albaladejo:2016hae}.

Extra information about the $D^*_{s0}(2317)$ can be obtained from the $B^+ \to \bar D^0 D^*_{s0}(2317)^+; D^*_{s0} \to D_s^+ \pi^0$ and $B^0 \to D^- D^*_{s0}(2317)^+; D^{*+}_{s0} \to D_s^+ \pi^0$ decays \cite{ParticleDataGroup:2024cfk,Belle:2015glz,BaBar:2004yux}.
The PDG compilation quotes the branching ratios~\cite{ParticleDataGroup:2024cfk}
\begin{align}\label{eq:Br1}
	&\mathcal{B}[B^+ \to \bar D^0 D^*_{s0}(2317)^+; D^{*+}_{s0} \to D_s^+ \pi^0]\nonumber \\
	=&0.79^{+0.15}_{-0.13} \times 10^{-3},
\end{align}
\begin{align}\label{eq:Br2}
	&\mathcal{B}[B^0 \to D^- D^*_{s0}(2317)^+; D^{*+}_{s0} \to D_s^+ \pi^0]\nonumber \\
	=&(1.05 \pm 0.16) \times 10^{-3}.
\end{align}
The reactions have received some theoretical attention from the perspective that the $D^*_{s0}(2317)$ is a $\bar s q$ state, using the factorization hypothesis \cite{LeYaouanc:2001ma,Cheng:2003kg,Cheng:2003id,Hsieh:2003xj,Datta:2003re,Datta:2004jx,Cheng:2003sm,Hwang:2004kga,Thomas:2005bu,Cheng:2006dm,Zhang:2021bcr}.
From the molecular perspective there is the work of Ref.~\cite{Faessler:2007cu}, where the $D^*_{s0}(2317)$ is assumed to be a molecule of $DK$.
The coupling of the resonance to the weak current, $W^\mu$, is constructed from the coupling of the resonance to the $DK$ pair and propagating the $DK$ to merge into $W^\mu$.
This requires a transition form factor $DKW^\mu$ which is taken from semileptonic decays studied in the literature.
For the other vertex of the $W^\mu$ one requires again the weak $BDW^\mu$ transition form factor, obtained from the literature, and finally the two blocks are multiplied, which is the basis of the factorization approximation.
The results obtained in Ref.~\cite{Faessler:2007cu} are in very good agreement with the results of Eqs.~\eqref{eq:Br1} and \eqref{eq:Br2}, coming from measurements done after the prediction of Ref.~\cite{Faessler:2007cu}.

Our approach to the problem also assumes the $D^*_{s0}(2317)$ to be of molecular nature, however coming from the $DK$ and $D_s\eta$ coupled channels, although the $DK$ channel is dominant.
However, in order to minimize uncertainties from the weak process and the use of the factorization approach, we follow a different strategy, and rely upon experimental data for $B\to \bar D DK$ production, from where we produce $\bar D D^*_{s0}(2317)$ by allowing the $DK$ component to propagate and merge into the $D^*_{s0}(2317)$.
The needed theoretical information is obtained from Refs.~\cite{Ikeno:2023ojl,Su:2025aiz}, and the information on the weak interaction is encoded in the $B\to \bar D DK$ experimental branching fractions.
This allows us to concentrate on the strong interaction producing the $D^*_{s0}(2317)$ in a relatively model independent way, which is most desirable to investigate the molecular structure of the $D^*_{s0}(2317)$ state.

\section{Formalism}
The first thing we wish to do is to express the branching ratios of Eqs.~\eqref{eq:Br1} and \eqref{eq:Br2} with the same normalization.
For this we consider that
\begin{equation}\label{eq:Frac1}
	\dfrac{\Gamma_{B^0}}{\Gamma_{B^+}}=1.08,
\end{equation}
and refer all decay widths to the one of $B^+$. 
Then Eq.~\eqref{eq:Br2} becomes
\begin{align}\label{eq:Br3}
	&\dfrac{\Gamma [B^0 \to D^- D^*_{s0}(2317)^+; D^{*+}_{s0} \to D_s^+ \pi^0]}{\Gamma_{B^+}} \nonumber \\[1.5mm]
	&=(1.13\pm 0.17) \times 10^{-3}. 
\end{align}
The rates of Eqs.~\eqref{eq:Br1} and \eqref{eq:Br3} are barely compatible within errors, and in our approach they should be practically equal, as we shall see.
For consistency with our approach we take a unique branching ratio equal to the average of the two rates summing errors in quadrature, which makes it compatible with the centroid of each branching ratio,
\begin{align}\label{eq:Br4}
	&\dfrac{\Gamma [B \to \bar D D^*_{s0}(2317)^+; D^{*+}_{s0} \to D_s^+ \pi^0]}{\Gamma_{B^+}} \nonumber \\[1.5mm]
	&=(0.96\pm 0.23) \times 10^{-3}.
\end{align}

\begin{figure}[b]
	\begin{center}
		\includegraphics[width=0.7\linewidth]{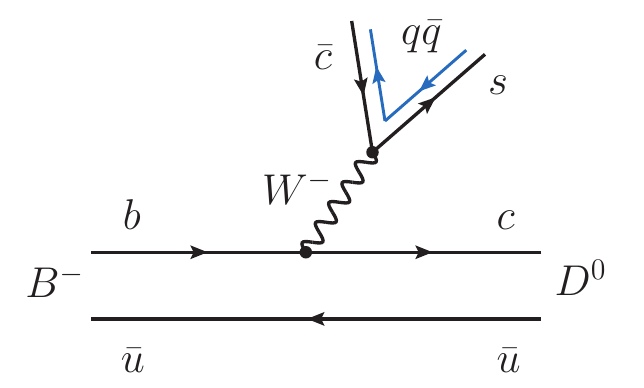}
		\vspace{-0.25cm}
		\caption{$B^-$ decay with external emission and hadronization of the $\bar c s$ quark pair.}
		\label{fig:Fig1}
	\end{center}
\end{figure}
\begin{figure}[b]
	\begin{center}
		\includegraphics[width=0.7\linewidth]{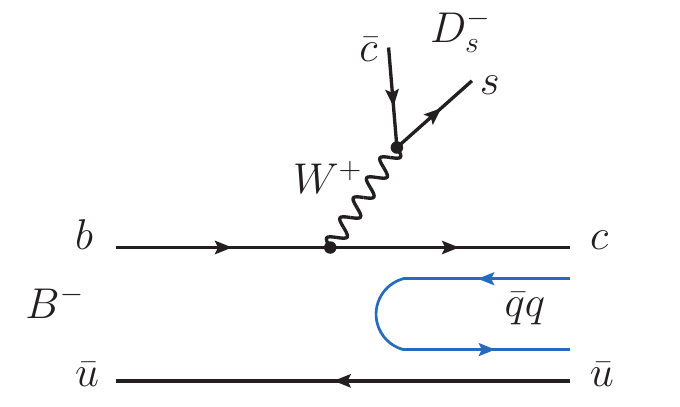}
		\vspace{-0.25cm}
		\caption{External emission for $B^-$ decay with hadronization of $c\bar u$ quark pair.}
		\label{fig:Fig2}
	\end{center}
\end{figure} 

Next we study the basic decay $B\to \bar D DK, \bar D D_s \eta$ since we need the $DK$ and $D_s \eta$ channels to build the $D^*_{s0}(2317)$ in the molecular picture.
For this, we look at the dominant decay modes, external and internal emission \cite{Chau:1982da} at the quark level and hadronize a $q\bar q$ component to have three mesons at the end.
First we look at the diagram of Fig.~\ref{fig:Fig1} for $B^-$ decay with external emission and Cabibbo allowed decay.
The hadronization of the $\bar cs$ pair proceeds as follows
\begin{equation}
  s\bar c \to \sum_i s\;\bar q_i q_i\; \bar c=\sum_i \mathcal{P}_{3i}\, \mathcal{P}_{i4}=(\mathcal{P}^2)_{34},
\end{equation}
where $\mathcal{P}$ is the $q\bar q$ matrix, most conveniently written in terms of pseudoscalar mesons as
\begin{equation}\label{eq:Pmatrix}
    \mathcal{P} =\scalebox{0.7}{
        $
        \left(
        \begin{array}{cccc}
            \frac{1}{\sqrt{2}}\pi^0 + \frac{1}{\sqrt{3}}\eta + \frac{1}{\sqrt{6}}\eta^{\prime} & \pi^+ & K^+  &  \bar D^0\\[2mm]
            \pi^- & -\frac{1}{\sqrt{2}}\pi^0 + \frac{1}{\sqrt{3}}\eta + \frac{1}{\sqrt{6}}\eta^{\prime} & K^0  & D^-\\[2mm]
            K^- & \bar{K}^0 & ~-\frac{1}{\sqrt{3}}\eta + \sqrt{\frac{2}{3}}\eta^{\prime}~  & D_s^-\\[2mm]
            D^0 & D^+ & D_s^+  & \eta_c\\
        \end{array}
        \right)$
	}.
\end{equation}
Then, neglecting $\eta'$ and $\eta_c$ terms which do not play a role here, we obtain 
\begin{equation}
  s\bar c \to K^- \bar D^0 + \bar K^0 D^- -\frac{1}{\sqrt{3}}\eta D_s^-.
\end{equation}
For the complex conjugate $B^+$ decay, we obtain
\begin{equation}\label{eq:HadB+}
  B^+ \to \bar D^0 \left(K^+ D^0 + K^0 D^+ -\frac{1}{\sqrt{3}}\,\eta D_s^+ \right).
\end{equation}

The same exercise with $B^0$ decay leads to 
\begin{equation}\label{eq:HadB0}
  B^0 \to D^- \left(K^+ D^0 + K^0 D^+ -\frac{1}{\sqrt{3}}\,\eta D_s^+ \right).
\end{equation}
This mechanism, factorizing the $B\to W^\mu D$ and $W^\mu \to DK$ transitions, is what is considered in Ref.~\cite{Faessler:2007cu}.
One can also hadronize the $c\bar u$ pair of Fig.~\ref{fig:Fig1} as shown in Fig.~\ref{fig:Fig2}.
We have now, 
\begin{align}\label{eq:cubar}
  c\bar u \to &\sum_i c\;\bar q_i q_i\; \bar u=\sum_i \mathcal{P}_{4i}\, \mathcal{P}_{i1}=(\mathcal{P}^2)_{41} \nonumber \\[1.5mm]
  &= D_s^- \left[ D^0 \left( \frac{\pi^0}{\sqrt{2}}+\frac{\eta}{\sqrt{3}}\right)+D^+ \pi^- +D_s^+ K^-\right],
\end{align}
and hence changing to $B^+$ decay
\begin{equation}\label{eq:HadB+2}
  B^+ \to D_s^+ \left[ \bar D^0 \left( \frac{\pi^0}{\sqrt{2}}+\frac{\eta}{\sqrt{3}}\right)+D^- \pi^+ +D_s^- K^+\right],
\end{equation}
and we see that only the $D_s^+\eta$ term contributes to the process that we are interested in. 
Similarly for $B^0$ decay, we would have an extra $D_s^+ \eta$ term from this mechanism. 
Now we have to hadronize the $c\bar d$ component in $\bar B^0$ decay, leading to
\begin{equation}\label{eq:12new}
  B^0 \to D_s^+ \left[ \bar D^0 \pi^- + D^-\left( -\frac{\pi^0}{\sqrt{2}}+\frac{\eta}{\sqrt{3}}\right)+D_s^- K^0\right].
\end{equation}

\begin{figure}[t]
	\begin{center}
		\includegraphics[width=0.7\linewidth]{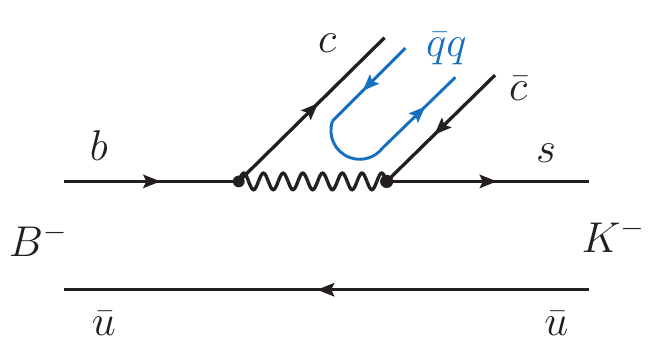}
		\vspace{-0.25cm}
		\caption{Internal emission for $B^-$ decay with $c\bar c$ hadronization.}
		\label{fig:Fig3}
	\end{center}
\end{figure}

Next we look at internal emission, a non factorizable contribution, which we depict in Fig.~\ref{fig:Fig3}.
Once again, the $c\bar c$ hadronization leads to
\begin{equation}
  c\bar c \to \sum_i c\;\bar q_i q_i\; \bar c=\sum_i \mathcal{P}_{4i}\, \mathcal{P}_{i4}=(\mathcal{P}^2)_{44},
\end{equation}
and hence
\begin{equation}\label{eq:HadB-}
  B^- \to K^- \left(D^0 \bar D^0 + D^+ D^- + D_s^+ D_s^- \right),
\end{equation}
or accordingly
\begin{equation}\label{eq:HadB+3}
  B^+ \to K^+ \left(D^0 \bar D^0 + D^+ D^- + D_s^+ D_s^- \right).
\end{equation}
Analogously we find
\begin{equation}\label{eq:HadB03}
  B^0 \to K^0 \left(D^0 \bar D^0 + D^+ D^- + D_s^+ D_s^- \right).
\end{equation}
We can see that Eqs.~\eqref{eq:HadB+3} and \eqref{eq:HadB03} provide the relevant terms $\bar D^0 K^+ D^0$ for $B^+$ decay and $D^- K^0 D^+$ in $B^0$ decay.

We give a weight $A$, to be obtained from experiment, to the mechanism of external emission of Eqs.~\eqref{eq:HadB+}, \eqref{eq:HadB0} for $B^+$ and $B^0$ decays, and a weight $A\beta$ for the terms of internal emission of Eqs.~\eqref{eq:HadB+3}, \eqref{eq:HadB03}. 
From the large $N_c$ counting we expect $\beta$ to be of the order of $1/3$.
We also give a different weight $A\gamma$ to the terms of Eqs.~\eqref{eq:HadB+2} and \eqref{eq:12new}, which correspond to the mechanism of Fig.~\ref{fig:Fig2}.
We thus conclude from our study that the internal plus external emission, followed by hadronization of some $\bar qq$ pair, leads to the hadronic components relevant to our processes
\begin{equation}\label{eq:HadB+4}
  B^+ \to A \, \bar D^0 \left[(1+\beta) K^+ D^0 + K^0 D^+ -\frac{1}{\sqrt{3}}(1-\gamma) \eta D_s^+ \right],
\end{equation}
\begin{equation}\label{eq:HadB04}
  B^0 \to A\, D^- \left[K^+ D^0 + (1+\beta) K^0 D^+ 
  -\frac{1}{\sqrt{3}}(1-\gamma) \eta D_s^+
  \right].
\end{equation}

At this point, it is worthwhile justifying the different weight of the mechanisms of Figs.~\ref{fig:Fig1} and \ref{fig:Fig2} (Eqs.~\eqref{eq:HadB+} \eqref{eq:HadB0} versus Eqs.~\eqref{eq:HadB+2} \eqref{eq:12new}). 
From Refs.~\cite{Gasser:1983yg,Scherer:2002tk}, one can see the structure of the $W^\mu$ leading to two pseudoscalar mesons, which goes as $W_\mu \langle [P, \partial_\mu P] T_-\rangle$, with $T_-$ a matrix related to the Cabibbo-Kobayashi-Maskawa elements (see Refs.~\cite{Ren:2015bsa,Sun:2015uva} for further details). 
With the dominant $\mu=0$ component, the $WPP$ vertex goes as $p_1^0 -p_2^0$, which vanishes in average when the two pseudoscalars have the same mass, which is not the case here.
But this particular structure does not appear in the mechanism of Fig.~\ref{fig:Fig2}, and taking effective weights for each of the mechanisms, the weights are certainly different, although we expect them to be of the same order of magnitude.

Next we would like to determine $A$ and $\beta$ from the experimental banching ratios for the $B^+ \to \bar D^0 K^+ D^0$, $B^+ \to \bar D^0 K^0 D^+$, $B^0 \to D^- K^+ D^0$ and $B^0 \to D^- K^0 D^+$ reactions \cite{ParticleDataGroup:2024cfk}.
Once again, we find opportune to elaborate on the origin of these rates.
The branching ratios for the last three reactions are obtained from a Babar experiment \cite{BaBar:2010tqo}.
The first one comes from averaging the results of the Babar experiment \cite{BaBar:2010tqo} with a Belle experiment \cite{Belle:2007hht}, which differ by nearly a factor of two.
Although the final results do not change much by taking the PDG average, we find more consistent to take the four rates from the same experiment \cite{BaBar:2010tqo}, and we have
\begin{align}
    \mathcal{B}[B^+ \to \bar D^0 K^+ D^0 ] &=(1.31\pm 0.14)\times 10^{-3}, \label{eq:BrDDK1} \\[2mm]
    \mathcal{B}[B^+ \to \bar D^0 K^0 D^+ ] &=(1.55\pm 0.21)\times 10^{-3}, \label{eq:BrDDK2} \\[2mm]
    \mathcal{B}[B^0 \to D^- K^+ D^0 ] &=(1.07\pm 0.11)\times 10^{-3}, \label{eq:BrDDK3} \\[2mm]
    \mathcal{B}[B^0 \to D^- K^0 D^+ ] &=(0.75\pm 0.17)\times 10^{-3}. \label{eq:BrDDK4}
\end{align}
The last two branching fractions, related to $\Gamma_{B^+}$ via Eq.~\eqref{eq:Frac1} give
\begin{align}
	&\mathcal{B}'[B^0 \to D^- K^+ D^0 ] =(1.16\pm 0.12)\times 10^{-3}, \label{eq:BrDDK5}\\[2mm]
	&\mathcal{B}'[B^0 \to D^- K^0 D^+ ] =(0.81\pm 0.18)\times 10^{-3}. \label{eq:BrDDK6}
\end{align}
According to Eqs.~\eqref{eq:HadB+4} and \eqref{eq:HadB04}, the $B^+ \to \bar D^0 K^0 D^+$ and $B^0 \to D^- K^+ D^0$ reactions proceed via external emission and should have the same decay width.
Similarly, the $B^+ \to \bar D^0 K^+ D^0$ and $B^0 \to D^- K^0 D^+$ reactions proceed via external and internal emission and should also have the same decay width.
This means that the rates of Eqs.~\eqref{eq:BrDDK2} and \eqref{eq:BrDDK5} should be equal, and so should be those of Eqs.~\eqref{eq:BrDDK1} and \eqref{eq:BrDDK6}. 
Actually they are compatible within errors.
To be consistent with our approach, we take the average of the rates with the error obtained summing the errors of each rate in quadrature. 
Then we take
\begin{align}
	\mathcal{B}[B^+ \to \bar D^0 K^0 D^+]=&\mathcal{B}'[B^0 \to D^- K^+ D^0 ] \nonumber\\
	=&(1.36\pm 0.24)\times 10^{-3}, \label{eq:BrDDK7}\\[2mm]
	\mathcal{B}[B^+ \to \bar D^0 K^+ D^0]=&\mathcal{B}'[B^0 \to D^- K^0 D^+ ] \nonumber\\
	=&(1.06\pm 0.23)\times 10^{-3}. \label{eq:BrDDK8}
\end{align}
From the former equations one might induce that $\beta$ is negative, with $\beta \simeq -0.1$ if we look only at the centroids of the data.
However, considering the errors, $\beta$ could also be of the order of $\beta\simeq 0.1$.
We can only conclude that $\beta$ is small, as expected, but in order to estimate uncertainties in the final results we will assume $\beta$ to run in a range of $\beta \in [-0.2, 0.2]$.

Next we determine $\frac{A^2}{\Gamma_{B^+}}$ using 
\begin{equation}\label{eq:AGam}
	\dfrac{1}{\Gamma_{B^+}}\, \dfrac{d\Gamma[B^+\to \bar D^0 K^0 D^+]}{dM_{\rm inv}(D^+ K^0)}=\dfrac{1}{(2\pi)^3}\; \dfrac{1}{4 M^2_{B^+}}\, p_{\bar D^0}\, \tilde{p}_{K^0}\; \dfrac{A^2}{\Gamma_{B^+}},
\end{equation}
with 
\begin{align}\label{eq:momen_p}
	p_{\bar D^0}=&\dfrac{\lambda^{1/2}(M^2_{B^+}, m^2_{\bar D^0}, M^2_{\rm inv}(D^+ K^0))}{2\, M_{B^+}},\\[2mm]
	\tilde{p}_{K^0}=&\dfrac{\lambda^{1/2}(M^2_{\rm inv}(D^+ K^0), m^2_{K^0}, m^2_{D^+})}{2\, M_{\rm inv}(D^+ K^0)}. 
\end{align}
Integrating Eq.~\eqref{eq:AGam} over $M_{\rm inv}(D^+ K^0)$ and equating the result to the rate of Eq.~\eqref{eq:BrDDK7}, we obtain $\frac{A^2}{\Gamma_{B^+}}$.

\begin{figure}[b]
	\begin{center}
		\includegraphics[width=0.8\linewidth]{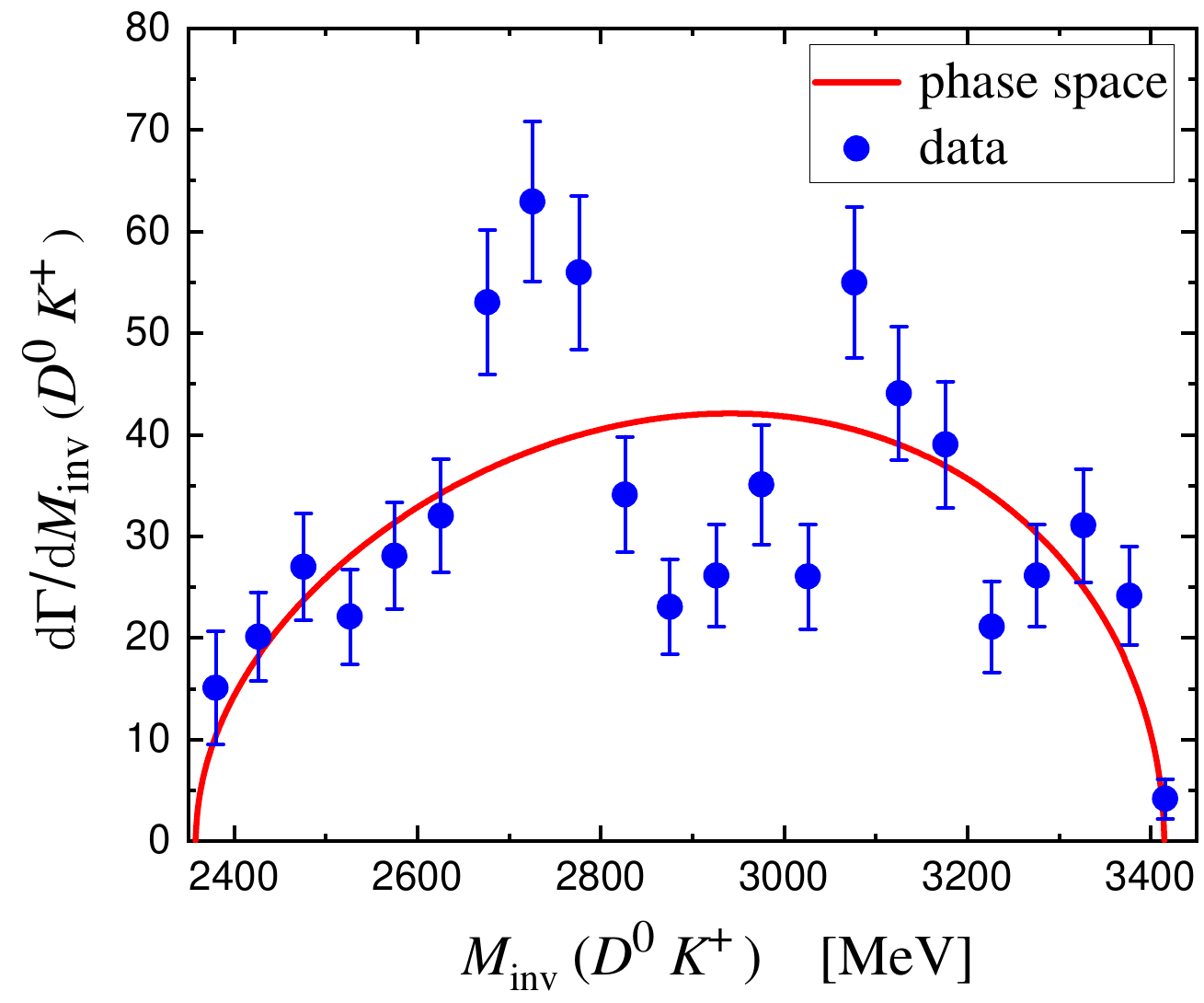}
		\vspace{-0.25cm}
		\caption{The $D^0K^+$ mass distribution for the $B^+\to \bar D^0 K^+ D^0$ reaction: data taken from Fig. 2(c) of Ref.~\cite{Belle:2007hht}; solid line the phase space, adjusted to reproduce the low mass region.}
		\label{fig:new}
	\end{center}
\end{figure}

The calculation of $\Gamma$ via Eq.~\eqref{eq:AGam} assumes that the $D^+ K^0$ mass distribution just follows phase space and proceeds via $S$-wave.
The transition matrix is given by the constant $A$.
This is certainly a simplification since there can be final state interaction and excitation of resonances, even $D$-wave contributions given the large phase space for the reactions.
For the reaction of Eq.~\eqref{eq:BrDDK1}, there is information on the mass distribution \cite{Belle:2007hht}, which we reproduce in Fig.~\ref{fig:new}.

In Fig.~\ref{fig:new}, we see that there are clear structures beyond the phase space.
According to Ref.~\cite{Belle:2007hht}, they correspond to reflections of the $\psi(3770)$ and $\psi(4160)$ excitation, decaying to $D^0 \bar D^0$.
In Ref.~\cite{Belle:2007hht}, one further shows spectra with cuts and further background subtraction. 
We are not interested in this detail. 
We are just interested in two aspects of the reaction: a) that we obtain a fair description of the integrated mass distribution, and b) that the transition amplitude that we use is fair in the low $D^0K^+$ region, since we shall use it to infer the excitation of the $D^*_{s0}(2317)$ which appears around $44 \mev$ below the $D^0K^+$ threshold.
In this latter case, we are interested in getting fair results in the lowest part of the $D^0K^+$ threshold.
We accomplish that using a constant transition amplitude, as one can see in Fig.~\ref{fig:new}, and then, without pretending to obtain the structure of the $D^0K^+$ distribution, what we see is that we obtain a fair integrated area, meaning that we can obtain a good mass distribution at lower $D^0K^+$ invariant mass distribution and a fair total width for $B^+\to \bar D^0 K^+ D^0$ reaction.

At this point, it is worth mentioning that we should expect the same mass distribution for the $B^0 \to D^- K^0 D^+$ reaction since the $\psi(3770)$ and $\psi(4160)$ resonances, having isospin $I=0$, will decay equally to $D^0\bar D^0$ and $D^+ D^-$.

The $D^0 K^+$ or $D^+ K^0$ interaction in these two reactions are also the same in $I=0$, the channel that has a strong interaction leading to the formation of the $D_{s0}^*(2317)$ resonance. 
However, the narrow $D_{s0}^*(2317)$ resonance appears about $44\mev$ below threshold and does not influence much the mass distributions of the $B^+\to \bar D^0 K^+ D^0$ and $B^0 \to D^- K^0 D^+$ reactions.
This discussion is made to justify our claim that the two reactions should have the same branching ratio.

In the $DK$ mass distributions of the $B^+\to \bar D^0 K^0 D^+$ and $B^0\to D^- K^+ D^0$ reactions with $D \bar D$ in $I=1$, one does not excite the $\psi$ resonances, and the $D^+ K^0$ and $D^0 K^+$ interactions would also be the same in the strong $I=0$ attractive channel, but as we mentioned above, the most important effect is the generation of the $D_{s0}^*(2317)$ around $44\mev$ below threshold, which does not influence much the $D^+ K^0$ or $D^0 K^+$ mass distributions.
We, hence, expect the mass distributions in these two cases to be very similar and justify that the rates should be the same for the two reactions, as we have assumed.

We should also mention that the $D_s^+ \eta$ term in Eqs.~\eqref{eq:HadB+4} and \eqref{eq:HadB04} does not play a role in any of the four reactions of Eqs.~\eqref{eq:BrDDK1}-\eqref{eq:BrDDK4}.
It could indirectly contribute via $D_s^+ \eta \to D^0 K^+ (D^+ K^0)$ through final state interaction, but this, again, would go in the region of the $D_{s0}^*(2317)$, far below the $DK$ threshold, and it is, hence negligible. 
However, this term will contribute when we produce the $D_{s0}^*(2317)$ resonance, through the reactions of Eqs.~\eqref{eq:Br1} and \eqref{eq:Br2}, which we discuss below.

\subsection{$B^+\to \bar D^0 D^*_{s0}(2317)^+$ and $B^0\to D^- D^*_{s0}(2317)^+$ decays}

\begin{figure*}[t]
	\begin{center}
		\includegraphics[width=0.68\linewidth]{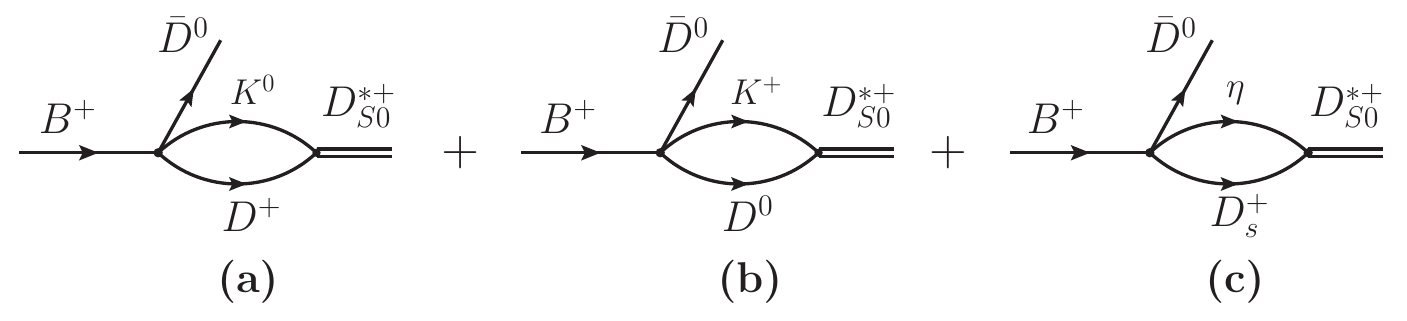}
		\vspace{-0.25cm}
		\caption{Final state interaction for $B^+\to \bar D^0 D^{*+}_{s0}$ through rescattering of the $DK, D_s\eta$ components.}
		\label{fig:Fig4}
	\end{center}
\end{figure*}

We assume that the $D^*_{s0}(2317)^+$ strong decay to $D_s^+ \pi^0$ basically exhausts all the $D^*_{s0}$ width, according to the BESIII work of Ref.~\cite{BESIII:2017vdm}, and evaluate the rates for $B^+\to \bar D^0 D^*_{s0}(2317)^+$ and $B^0\to D^- D^*_{s0}(2317)^+$, taking into account the interaction of the $DK$ components, as shown diagrammatically in Fig.~\ref{fig:Fig4}.

The transition matrix $t$ for the diagrams of Fig.~\ref{fig:Fig4} is given by
\begin{align}\label{eq:tB+}
   t_{B^+}&=A\Big[ G_{D^+K^0}(M_{D^*_{s0}})\, g_{D^+K^0}  \nonumber\\[1.5mm]
   &~~~~~ + (1+\beta)\, G_{D^0 K^+} (M_{D^*_{s0}})\, g_{D^0 K^+} \nonumber\\[1.5mm]
   &~~~~~ -\dfrac{1}{\sqrt{3}} (1-\gamma)\, G_{D_s^+ \eta} (M_{D^*_{s0}})\, g_{D_s^+ \eta}\Big],
\end{align}	
where $G_{D^+K^0}$, $G_{D^0K^+}$ are the loop functions of $D^+K^0$, $D^0K^+$ respectively and $g_{D^+K^0}$, $g_{D^0K^+}$, $g_{D_s^+ \eta}$ are the couplings of the $D^{*+}_{s0}$ resonance to the $D^+ K^0$, $D^0 K^+$, and $D_s^+ \eta$ components respectively.
Similarly, the amplitude for $B^0 \to D^- D^*_{s0}(2317)$ decay is given by
\begin{align}\label{eq:tB0}
   t_{B^0}&=A\Big[ (1+\beta)\,G_{D^+K^0}(M_{D^*_{s0}})\, g_{D^+K^0}  \nonumber\\[1.5mm]
   &~~~~~~ +  G_{D^0 K^+} (M_{D^*_{s0}})\, g_{D^0 K^+}\nonumber\\[1.5mm]
   &~~~~~~-\dfrac{1}{\sqrt{3}} (1-\gamma)\, G_{D_s^+ \eta} (M_{D^*_{s0}})\, g_{D_s^+ \eta}\Big].
\end{align}	
The couplings $g_{D^+K^0}$, $g_{D^0K^+}$ and $g_{D_s^+ \eta}$ are obtained in the study of the strong and radiative decays of the $D^*_{s0}(2317)$, and are given by \cite{Su:2025aiz}
\begin{equation}
	\begin{aligned}\label{eq:couplings}
g_{D^+K^0}&=8252 -i69 \mev,\\[1.5mm]
g_{D^0K^+}&= 8129+i76\mev,\\[1.5mm]
g_{D_s^+ \eta}&= -6312\mev,
	\end{aligned}
\end{equation}
which are practically equal, guaranteeing that the rates for $B^+$ or $B^0$ decay into the $D^*_{s0}(2317)$ are basically equal.
The difference of the $DK$ couplings is small, of the order of $1.5\%$.
If they were exactly equal, and we considered also equal $K^+, K^0$ masses in the $G$ loops, Eqs.~\eqref{eq:tB+} and \eqref{eq:tB0} would lead to the same rate.
In the presence of the internal emission and different couplings, one can have a small difference in the production rates of the $D^*_{s0}(2317)$ in $B^+$ or $B^0$ decays.
Yet, in view of the  small difference in the couplings compared to the experimental errors, the small differences in the $G$ loops from having different $K^+, K^0$ masses (which is relevant for isospin violating processes, but not here), and the ignorance of the value of the $\beta$ parameter of internal emission, we consider a unique coupling, as average of those in Eq.~\eqref{eq:couplings}.
We take
\begin{equation}
   g_{DK}=8191 \mev.
\end{equation}	

In order to estimate uncertainties in our results, we consider also the couplings obtained in Ref.~\cite{Li:2024rlw} from the study of the $(\bar D \bar K)$ mass distributions in the $\Lambda_b \to \Lambda_c (\bar D \bar K)^-$ decays,
	\begin{align}\label{eq:couplings-2}
g_{D^+ K^0}&=8145 \mev, \nonumber\\[1.5mm]
g_{D^0 K^+}&=8182 \mev,\\[1.5mm]
g_{D_s^+ \eta}&= -5571 \mev. \nonumber
	\end{align}
These values are very similar to those of Eq.~\eqref{eq:couplings}. 
As mentioned above, we take now an averaged value of the four $g_{DK}$ coupling constants from Eqs.~\eqref{eq:couplings} and \eqref{eq:couplings-2} and also the average value of $g_{D_s^+ \eta}$ and have 
\begin{equation}
	\begin{aligned}\label{eq:couplings-3}
g_{DK}&=8187 \mev,\\[1.5mm]
g_{D_s^+ \eta}&= -5942 \mev.
	\end{aligned}
\end{equation}
with uncertainties of about $1\%$ in $g_{DK}$ and $6\%$ in $g_{D_s^+ \eta }$.

The decay width is given by
\begin{equation}\label{eq:Gamma_i}
   \Gamma_i =\dfrac{1}{8\pi}\; \dfrac{1}{M^2_{B_i}}\; |t_i|^2\, q_i,
\end{equation}
with
\begin{equation}
	q_i=\dfrac{\lambda^{1/2}(M^2_{B_i}, m^2_{\bar D_i}, M^2_{D^*_{s0}})}{2\, M_{B_i}}.
\end{equation}
We apply it to the $B^+ \to \bar D^0 D^{*}(2317)^+$ decay, which serves us for $B^0 \to D^- D^{*}(2317)^+$ ignoring small differences in the masses of $\bar D_i$.
The strategy to calculate $\Gamma$ and the error consists in taking random numbers of the branching fraction of $B^+ \to \bar D^0 K^0 D^+$ (Eq.~\eqref{eq:BrDDK7}) within the error band,
to get $A^2/\Gamma_{B^+}$ according to Eq.~\eqref{eq:AGam}, within errors. 
Then we fit the data of Eq.~\eqref{eq:Br4} with the formula of Eq.~\eqref{eq:Gamma_i} and the couplings of Eq.~\eqref{eq:couplings-3}, together with $\beta$ in the range of $\beta \in [-0.2, 0.2]$, to obtain the parameter $\gamma$ within errors.

\section{Results}
We start from the data of Eq.~\eqref{eq:BrDDK7} and generate random numbers with a Gaussian distribution according to the centroid and the error of the branching fraction.
By using Eq.~\eqref{eq:AGam}, we evaluate the value of $A'\equiv A^2/\Gamma_{B^+}$, 
and from the results we take the average, $\bar A'$, and the  dispersion, taking $(\Delta A')^2= \frac{1}{N}\sum_i (A'_i-\bar A')^2$, with $N$ the number of random points. 
We obtain
\begin{align}\label{eq:ResA}
	\dfrac{|A|^2}{\Gamma_{B^+}}=&0.056 \pm 0.008 {\mev}^{-1}.
\end{align}

Then we fit the central value of the data of Eq.~\eqref{eq:Br4} with the formula of Eq.~\eqref{eq:Gamma_i} and the couplings of Eq.~\eqref{eq:couplings-3},  together with $\beta=0$, to get the central value of the parameter $\gamma$, which we find to be $\gamma=-1.57$\footnote{Mathematically we also obtain a solution with $\gamma \sim 20$, but since the mechanisms of Figs.~\ref{fig:Fig1} and \ref{fig:Fig2} should have similar weights, we discard this solution as unphysical.}, with the order of magnitude of $1$ as we expected.

As a next step, according to Eq.~\eqref{eq:Gamma_i}, we use the related parameters $A^2/\Gamma_{B^+}$, $\beta$, $\gamma$ and the couplings $g_{DK},~ g_{D_s^+ \eta}$, with their uncertainties, to evaluate the branching fraction of $\Gamma [B \to \bar D D^*_{s0}(2317)]/\Gamma_{B^+}$ within errors.
As mentioned above, we take the parameter $\beta$ in the range of 
\begin{equation}
	\beta \in [-0.2, 0.2].
\end{equation}
We find that the results are relatively insensitive to the precise value of $\gamma$. 
In fact, we tune the error in $\gamma$ to find the errors in Eq.~\eqref{eq:Br4}, and we find this error to be of the order of $100\%$,
\begin{equation}
	\gamma=-1.57\pm 1.50.
\end{equation}
We generate simultaneously random numbers for $A^2/\Gamma_{B^+}$, $\beta$, $\gamma$, $g_{DK}$ and $g_{D_s^+ \eta}$ with Gaussian distributions according to the centroids and errors obtained above, and evaluate $\Gamma [B \to \bar D D^*_{s0}(2317)]/\Gamma_{B^+}$ with Eq.~\eqref{eq:Gamma_i} in each random run.
The result obtained is the following:
\begin{align}\label{eq:BrTheo}
	&\dfrac{\Gamma [B \to \bar D D^*_{s0}(2317)^+; D^{*+}_{s0} \to D_s^+ \pi^0]}{\Gamma_{B^+}}\Big|_{\rm theo.}\nonumber\\
	&=(0.92\pm 0.22) \times 10^{-3},
\end{align}
which is compatible with the data of Eq.~\eqref{eq:Br4}.

According to Eqs.~\eqref{eq:tB+} and \eqref{eq:tB0}, the parameter $\gamma$ is related solely to the $D_s^+ \eta$ channel and does not affect the rates in Eqs.~\eqref{eq:BrDDK1}-\eqref{eq:BrDDK4}, as these rates are evaluated at the tree level.  
The $D_s^+ \eta$ channel only becomes relevant in the $B\to \bar D D_{s0}^*(2317)$ reactions where the $D_{s0}^*(2317)$ resonance is produced via the rescattering mechanism illustrated in Fig.~\ref{fig:Fig4}.  
As can be seen from the couplings of the resonance to the $D_s^+ \eta$ channel in Eqs.~\eqref{eq:couplings} and \eqref{eq:couplings-2}, which are smaller than those to the $DK$ channels, and given that the $D_s^+ \eta$ channel lies further away from the $D_{s0}^*(2317)$ mass, the relative importance of the $D_s^+ \eta$ channel is far smaller than that of the $DK$ channels.
Actually, if we neglect the $D_s^+ \eta$ channel, we obtain a branching ratio for $B\to \bar D D_{s0}^*(2317)$ of the order of $(0.6\pm 0.1) \times 10^{-3}$, which, within errors, is compatible with the value $(0.96\pm 0.23)\times 10^{-3}$ of Eq.~\eqref{eq:Br4}.
The consideration of the $D_s^+ \eta$ channel allows one to get a better solution.

All these considerations tell us that the values obtained from our analyses are consistent with the molecular picture of the $D_{s0}^*(2317)$ state, but one should not forget that the resonance is not $100\%$ molecular.
Indeed, as found in Ref.~\cite{MartinezTorres:2014kpc}, the molecular probability of the $D_{s0}^*(2317)$ state is $(72\pm 13 \pm 5)\%$, indicating that there could still be some small contribution to the $B\to \bar D D_{s0}^*(2317)$ reaction from some non-molecular component of the $D_{s0}^*(2317)$.
With all these considerations and the results obtained here, one can conclude that the mostly molecular picture of the $D_{s0}^*(2317)$ is compatible with the four $B$ decays of Eqs.~\eqref{eq:BrDDK1}-\eqref{eq:BrDDK4} and the two $B$ decay ratios of Eqs.~\eqref{eq:Br1} and \eqref{eq:Br2}, without going to the extreme that these reactions are a proof to the molecular nature of the resonance.
Ultimately, it is the piling up of consistency with many reactions what gives support to one or another picture, and the present reactions should be looked at in this context.

\section{Conclusions}
We have studied the $B^{+(0)} \to \bar D^{0(-)} D^{*}_{s0}(2317)^+$ reactions from the perspective that the $D^*_{s0}(2317)$ resonance is a molecular state of the $DK, D_s \eta$ components. 
The strategy followed to study these processes differs from other works that rely upon the factorization method, taking information from semileptonic decays. We instead separate the weak decay process from the strong interaction that is responsible for the formation of the $D^*_{s0}$ from its molecular components.  
For this purpose we look at the $B^+\to \bar D^0 K^+ D^0$, $B^+\to \bar D^0 K^0 D^+$, $B^0 \to D^- K^+ D^0$, $B^0 \to D^- K^0 D^+$ reactions, which produce the $D^0 K^+$ and $D^+ K^0$ pairs, together with a $\bar D$, and use their experimental branching ratios, which encode the dynamics of the weak interaction. 
After this, the $D^0 K^+$, $D^+ K^0$ and $D_s^+ \eta$ components are allowed to propagate and interact, fusing into the $D^*_{s0}(2317)$ resonance, through couplings that can be evaluated theoretically from the strong interaction of these pseudoscalar mesons. 

We also look at the $B^+\to \bar D^0 K^+ D^0$, $B^+\to \bar D^0 K^0 D^+$, $B^0 \to D^- K^+ D^0$, $B^0 \to D^- K^0 D^+$ reactions from a microscopical point of view, showing that they can be formed from external and internal emission, followed by hadronization of $q\bar q$ pairs. 
In this way, internal emission, which is not factorizable, is also taken into account. 
     
By means of one free parameter, we could obtain a fair description of the $B^+\to \bar D^0 K^+ D^0$, $B^+ \to \bar D^0 K^0 D^+$, $B^0 \to D^- K^+ D^0$ and $B^0 \to D^- K^0 D^+$ reactions, together with the $B^+ \to \bar D^0 D_{s0}^*(2317)$ and $B^0 \to D^- D_{s0}^*(2317)$ reactions.
The agreement was better introducing the $D_s^+ \eta$ channel, which is only operative in the last two reactions, together with a new free parameter.
The analysis, short of a proof of the molecular picture of the $D_{s0}^*(2317)$ state, has to be considered as a test of consistency of the picture, which together with other tests in different reactions, like the $(\bar D \bar K)$ mass distribution in $\Lambda_b \to \Lambda_c (\bar D \bar K)$ \cite{Li:2024rlw}, the $DK$ mass distribution in $\bar B^0_s \to D_s^- (DK)^+$ \cite{ Albaladejo:2015kea}, the rate for $\bar B^0_s \to D_{s0}^*(2317) \bar \nu_e e^-$ and the $DK$ mass distribution in $\bar B^0_s \to DK \bar \nu_e e^-$ \cite{Navarra:2015iea}, or the mass distributions in $B_s \to \pi^+ \bar D^0 K^-$ \cite{Albaladejo:2016hae}, should serve to provide further support for this picture.

\section*{Acknowledgments}
This work is partly supported by the National Natural Science Foundation of China (NSFC) under Grants No. 12575081 and No. 12365019,
and by the Natural Science Foundation of Guangxi province under Grant No. 2023JJA110076,
and by the Central Government Guidance Funds for Local Scientific and Technological Development, China (No. Guike ZY22096024).
This work is also partly supported by the Spanish Ministerio de Economia y Competitividad (MINECO) and European FEDER funds under Contracts No. FIS2017-84038-C2-1-PB, PID2020-112777GB-I00, and by Generalitat Valenciana under contract PROMETEO/2020/023. 
This project has received funding from the European Union Horizon 2020 research and innovation program under the program H2020-INFRAIA-2018-1, grant agreement No. 824093 of the STRONG-2020 project.

\bibliographystyle{a}
\bibliography{refs}
\end{document}